\documentclass[3p,times,procedia]{elsarticle}
\usepackage{nupha_ecrc}

\volume{00}

\firstpage{1}

\journalname{Nuclear Physics A}

\runauth{E.~Iancu et al.}

\jid{nupha}

\jnltitlelogo{Nuclear Physics A}

\usepackage{amssymb}

\usepackage{bm,amsthm}

\usepackage[figuresright]{rotating}

\long\def\comment#1{ }
\newcommand{\tqq}{\theta_{q\bar q}}
\newcommand{\tf}{t_{\rm f}}
\newcommand{\abar}{\bar{\alpha}_s}

\newcommand{\beq}{\begin{equation}}
\newcommand{\eeq}{\end{equation}}

\begin{document}

\begin{frontmatter}



\dochead{XXVIIth International Conference on Ultrarelativistic Nucleus-Nucleus Collisions\\ (Quark Matter 2018)}

\title{Adding vacuum branching to jet evolution in a dense medium}

\author[sac]{P. Caucal}

\author[sac]{E.~Iancu}


\author[col]{A.H.~Mueller}

\author[sac]{G. Soyez}

\address[sac]{Institut de physique th\'{e}orique, Universit\'{e} Paris Saclay, CNRS, CEA, F-91191 Gif-sur-Yvette, France}

\address[col]{Department of Physics, Columbia University, New York, NY 10027, USA}

\begin{abstract}
 We study the fragmentation of a jet propagating in a dense  quark-gluon plasma. We show that the ``vacuum-like'' emissions triggered by the parton  virtualities can be factorized from the medium-induced radiation responsible for the energy loss within a controlled, ``double-logarithmic'', approximation in perturbative QCD. We show that the collisions with the plasma constituents modify the vacuum-like parton shower already at leading twist, in two ways: the radiation phase-space is reduced and the first emission outside the medium can violate angular ordering.  We compute the jet fragmentation function and find results in qualitative agreement with measurements at the LHC. 

\end{abstract}

\begin{keyword}
Quark Gluon Plasma; Jets; Jet quenching; Bremsstrahlung.

\end{keyword}

\end{frontmatter}




 \section{Introduction}\label{}


The energy lost by a jet propagating through the dense environment of an ultrarelativistic heavy ion collision and, more generally, the medium-induced modifications in the jet branching pattern, are among the main observables used in the study of the quark-gluon plasma (QGP) expected to be created in the intermediate stages of such a collision. For the respective data at RHIC and the LHC to be fruitfully exploited, it is essential to have a good theoretical understanding of the interactions between the jet and the QGP, from first principles. It is in particular of utmost importance to understand how the physics of parton branchings is modified by these interactions. On conceptual grounds, it is quite clear that the overall jet structure should get built via the interplay between two mechanisms for radiation: the usual, ``vacuum-like'', bremsstrahlung through which a virtual parton evacuate his virtuality (until this becomes as small as the hadronisation scale) and the additional, ``medium-induced'', radiation, which is triggered by the collisions between the partons from the jet and those from the plasma. Taken separately, these two mechanisms are by now rather well understood (notably due to recent progress with understanding color decoherence \cite{MehtarTani:2010ma,CasalderreySolana:2011rz,CasalderreySolana:2012ef} and multiple medium-induced branchings \cite{Blaizot:2012fh,Blaizot:2013hx,Kurkela:2014tla,Blaizot:2014ula,Iancu:2015uja,Escobedo:2016jbm}),  but it appears as a challenge to construct a unified theoretical picture which consistently encompasses both sources of radiation and their potential interplay. There are several reasons for such a difficulty: the respective underlying mechanisms are characterized  by different evolution variables --- parton virtualities for the vacuum-like emissions (VLEs), respectively, in-medium propagation time for the medium-induced ones (MIEs) ---, by different coherence properties (see below), and by different splitting functions --- the DGLAP splitting functions for the VLEs, respectively the BDMPS-Z spectrum (playing the role of a branching rate  \cite{Blaizot:2012fh,Blaizot:2013hx}) for the MIEs \cite{Baier:1996kr,Zakharov:1996fv,Wiedemann:2000za}. In the recent paper \cite{Caucal:2018dla}, we made a significant step towards such a unified description, by showing that the two mechanisms can be factorized from each other within controlled approximations in perturbative QCD.

 \begin{figure}
\centerline{
\includegraphics[width=.52\textwidth]{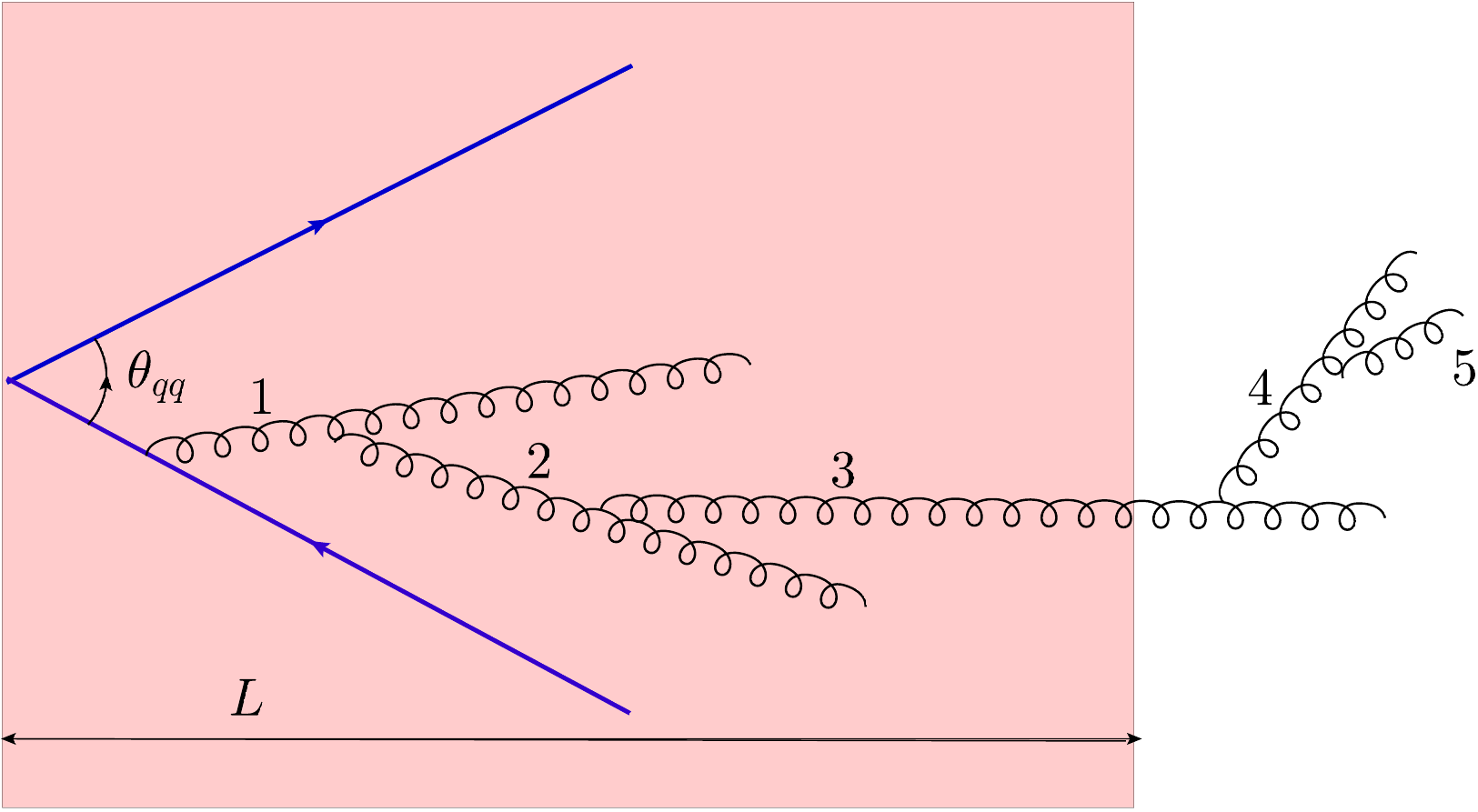}\qquad
\includegraphics[width=.4\textwidth]{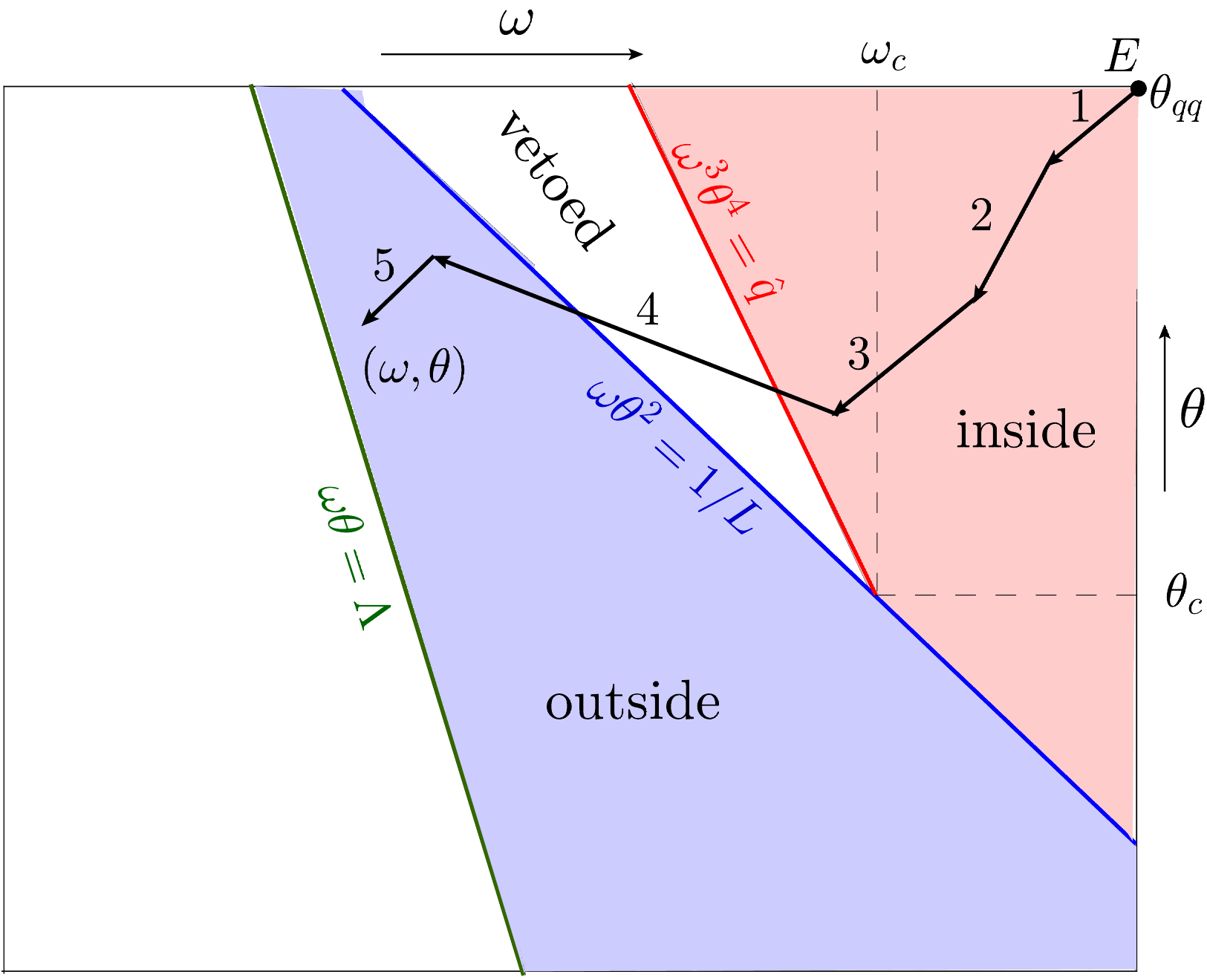}
}
\caption{\small Left: Pictorial representation of a typical vacuum-like parton cascade. The first 3 emissions occur inside
the medium and the last 2 outside. Angular ordering is violated by the 4th emission, but preserved by all the others. Right: The $(\omega, \,\theta)$ phase-space for VLEs in double-logarithmic units. The ``vetoed'' region is delimitated by the lines $\tf =L$ and $\tf=\sqrt{2\omega/\hat q}$
with $\tf\equiv 2/({\omega\theta^2})$. The branching process stops at the hadronization line $\omega\theta=\Lambda$.}
\label{figVLE}
\end{figure}

\section{Vacuum-like emissions in the double logarithmic approximation}

It is conceptually simpler to consider a jet initiated by a quark-antiquark antenna in a color singlet state with opening angle $\tqq < 1$, e.g. produced by the decay of a boosted $W/Z$ boson or a virtual photon. (For a generic jet which is produced by a parton, the role of $\tqq$ would be played by the jet radius $R$.) The quark and the antiquark are assumed to have equal energies: $E_q=E_{\bar q}\equiv E$, with $E$ a very high energy compared to the typical scales of the medium (see below). Also, the antenna is assumed to be produced directly inside the medium (a static and uniform plasma characterized for the present purposes by the jet quenching parameter $\hat q$) and to cross the medium along a distance $L$. Finally, for simplicity we shall work in the limit of a large number of colors $N_c\gg 1$, where a gluon emission can be pictured as the splitting of one dipole into two.
The quark and antiquark legs of this antenna have a large virtuality $Q^2\simeq (E\tqq)^2$, so they can radiate gluons like in the vacuum, i.e. via bremsstrahlung (see the left Figure \ref{figVLE} for an illustration). 

{\it The formation time.} Consider first a VLE which occurs {\em inside} the medium, meaning that the respective formation time $\tf$ is smaller than $L$. $\tf$ is
determined by the condition that the transverse separation
$\Delta r\sim \theta\tf$ between the gluon and its parent parton at
the time of emission be as large as the gluon transverse wavelength
$2/k_\perp$, with $k_\perp\simeq \omega\theta$ its transverse momentum
w.r.t. its emitter. This argument applies to both
vacuum-like and medium-induced emissions and implies
$\tf\simeq {2}{\omega}/k_\perp^2 \simeq {2}/({\omega\theta^2})$.
Then, gluons emitted inside the medium have a minimum $k_\perp$ set by
the momentum acquired via multiple collisions during formation:
$k_{\rm f}^2\simeq \hat q\tf$. This translates into an upper limit
$\tf\lesssim \sqrt{2\omega/\hat q}$ on the formation time, leaving two
possibilities: \texttt{(a)} {MIEs}, for which
$k_\perp\simeq k_{\rm f}$, so the corresponding formation time
saturates the upper limit, and \texttt{(b)} {VLEs}, 
which are comparatively harder, $k_\perp\gg k_{\rm f}$, meaning that
  they occur much faster than MIEs with the same energy:
\begin{equation}\label{tfvac} 
  \tf=\frac{2}{\omega\theta^2}\ll\sqrt{\frac{2\omega}{\hat q}}
  \ \Leftrightarrow\ \
  \omega\gg \Big(\frac{2\hat q}{\theta^4}\Big)^{\!\frac{1}{3}}\!\equiv \omega_{0}(\theta).\qquad\mbox{(VLE)}
\end{equation}
This constraint, which can be also formulated as a lower limit 
on the emission angle, applies only so long $\sqrt{2\omega/\hat q}< L$, i.e. for energies
$\omega\le \omega_c\equiv \hat q L^2/2$. Emissions
with larger energies ($\omega\ge \omega_c$) behave exactly as in the
vacuum: their emission angle can be arbitrarily small and their
formation time can be larger than $L$. We shall assume that $E > \omega_c$, which is 
indeed the case for the high energy ($E> 100$~GeV) jets at the LHC.

{\it The vetoed region.}
Eq.~(\ref{tfvac}) immediately implies the existence of a {\em vetoed region} in the $(\omega, \,\theta)$ phase-space for VLEs \cite{Caucal:2018dla}:
for a given energy $\omega\le \omega_c$, there are no VLEs with formation time within the range $\sqrt{2\omega/\hat q}< \tf < L$. As visible in Fig.~\ref{figVLE} (right), this excluded region exists only for angles larger than a special value
$\theta_c\equiv 2/\sqrt{\hat q L^3}$, which is quite small, $\theta_c < 0.1$, for the phenomenologically relevant
values for $\hat q$ and $L$.

\begin{figure}[t] \centerline{
\includegraphics[width=0.48\columnwidth]{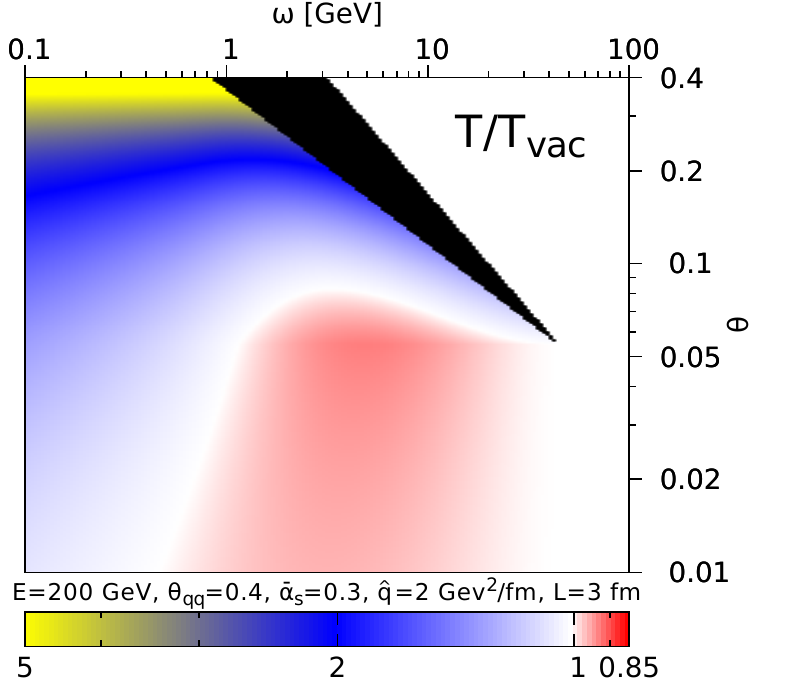}\qquad \includegraphics[width=0.5\columnwidth]{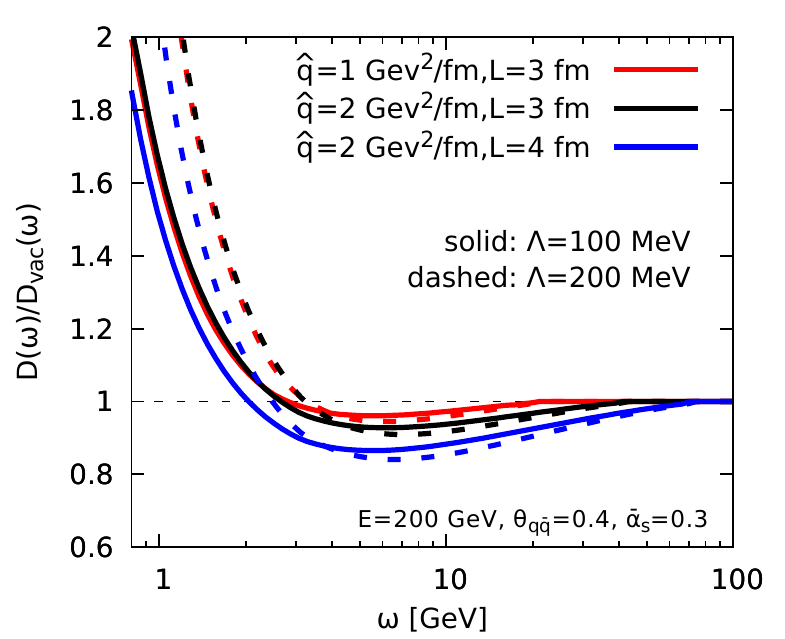}}
 \caption{\small Left: The ratio $T(\omega,\theta^2)/T_{\rm vac}(\omega,\theta^2)$ between the two-dimensional gluon distributions  in the medium and respectively the vacuum, both computed to DLA and for the values of the free parameters $E$, $\tqq$, $\abar$, $\hat q$ and $L$ shown in the figure.  Right: The ratio $D(\omega)/D_{\rm vac}(\omega)$ between the fragmentation functions  in the medium and respectively the vacuum, for different choices for the medium parameters $\hat q$ and $L$ and the hadronisation scale $\Lambda$.}
 \label{result-T}
\end{figure}

{\it Colour (de)coherence.}
For emissions by a colour-singlet antenna, even a VLE
obeying (\ref{tfvac}) could be still affected by the medium, via {\em color decoherence}
\cite{MehtarTani:2010ma,CasalderreySolana:2011rz,CasalderreySolana:2012ef}.
In the vacuum, gluon emissions at large angles $\theta\gg \tqq$ are
suppressed by the destructive interferences between the quark and the
antiquark. But an antenna propagating through a
dense QGP can lose its coherence via rescattering off
the medium: the quark and the antiquark suffer independent color
rotations, hence the probability that the antenna remains in a color
singlet state decreases with time.
The two legs of the antenna start behaving like independent
color sources after a time $t\sim t_{\rm coh}$, where $t_{\rm coh}\equiv\,\left({4}/{\hat
      q\tqq^2}\right)^{1/3}$ \cite{CasalderreySolana:2011rz}.
This scale would be comparable to $L$ for 
$\tqq\sim\theta_c$. In practice though, one generally has $\tqq\gg\theta_c$,
meaning that our original antenna will lose coherence pretty fast. In spite of that,
it cannot radiate vacuum-like  gluons at large angles $\theta\gg \tqq$, as 
noticed in \cite{Caucal:2018dla}. Indeed one can easily check that VLEs with
$\theta\gg \tqq$ would have formation times {\em even smaller} than the decoherence
time, ${2}/{\omega\theta^2}\ll  t_{\rm coh}(\tqq)$, hence they are killed by
interference, like in the vacuum. This implies that color decoherence plays no
special role for VLEs occurring inside the medium:
 only emissions with $\theta\lesssim\tqq$ are allowed whether
or not they occur at times larger than the decoherence time $t_{\rm coh}(\tqq)$.

{\it Multiple emissions inside the medium.} 
So far, we have considered a single emission inside the medium, but 
the previous arguments remain valid for an arbitrary numbers of successive VLE
which are strongly ordered in both energies and angles:
$\tqq\gg\theta_1\gg\cdots \gg\theta_n\gg\theta_c$ and
$E\gg\omega_1\gg\cdots\gg\omega_n\gg\omega_{0}(\theta_n)$.
These are precisely the cascades which give the dominant, double-logarithmic, contribution to the jet multiplicity
(or fragmentation function) $D(\omega)=\omega (d N/d\omega)$ when $\omega\ll E$ in pQCD. The fact
that the angular ordering of successive emissions can be preserved inside the medium (like in the vacuum) is highly non-trivial and follows from our previous observation that color decoherence via rescattering plays no role for the VLEs. 

{\it First emission outside the medium.}  The partons produced inside the medium via VLEs can act as sources for medium-induced radiation (and thus contribute to the energy loss by the jet), but they can also radiate gluons directly {\em outside} the medium. The first such an emission, whose formation time is, by definition, larger than $L$,  is rather special \cite{Caucal:2018dla}.  Indeed, all the in-medium sources with $\theta\gg\theta_c$ satisfy $t_{\rm coh}(\theta)\ll L$ and thus lose color coherence after propagating over a distance $L$ in the medium. Accordingly,  the first emission outside the medium can violate angular ordering. (A similar idea appears in~\cite{Mehtar-Tani:2014yea}.) This is important since it re-opens the angular phase-space for the subsequent cascades developing outside the medium (which are of course angular-ordered). A gluon cascade with these characteristics is illustrated in Fig.~\ref{figVLE} (left).

{\it Jet fragmentation function.} These physical considerations can be easily transposed into a calculation of the jet multiplicity in the double logarithmic approximation (DLA),  with the results shown in Fig.~\ref{result-T} \cite{Caucal:2018dla}. The left figure refers to the  two-dimensional gluon distribution $  T(\omega,\theta)\equiv \omega
  \theta^2({d^2 N}/{d\omega d\theta^2})$: we more precisely show the ratio
$T(\omega,\theta^2)/T_{\rm vac}(\omega,\theta^2)$ between the distribution generated 
in the presence of the medium and that in the vacuum.
This ratio is 1 for all the points either
inside the medium or with $\omega >\omega_c$. However, one sees
significant deviations from unity for points outside the medium with
energies $\omega < \omega_c$: for intermediate values of $\omega$ and
relatively small angles $\theta\lesssim 0.1\tqq$, one sees a small but
significative suppression compared to the vacuum (up to 15\%). For
smaller energies and larger angles, $\theta> 0.2$, one rather sees a
strong enhancement, owing to emissions violating angular ordering.

Given the two-dimensional distribution $  T(\omega,\theta)$, the fragmentation function $D(\omega)$ is obtained
by integrating over the angles. The right plot  in Fig.~\ref{result-T}  shows the ratio $D(\omega)/D_{\rm vac}(\omega)$. One sees a slight suppression (relative to vacuum) at intermediate energies, roughly
from 2~GeV up to  $\omega_c$, and a substantial
enhancement at lower energies $\omega \lesssim 2$~GeV. This enhancement is attributed 
to small-angle emissions inside the medium, radiating at larger angles outside the medium due to the lack
of angular ordering.
These results are in qualitative agreement with the respective
LHC measurements for the most central
PbPb collisions~\cite{Chatrchyan:2014ava,Aad:2014wha}. 

Since based on a probabilistic picture, our approach is suitable for Monte-Carlo implementations, which 
would allow to go beyond the present, double-logarithmic, approximation.

\smallskip
\noindent{\bf Acknowledgements} The work of E.I. and G.S. is supported in part by the Agence Nationale de la Recherche project  ANR-16-CE31-0019-01.   The work of A.H.M.
is supported in part by the U.S. Department of Energy Grant \# DE-FG02-92ER40699.








\end{document}